\documentclass{article}
\usepackage{amsmath}
\usepackage{jheppub}
\usepackage{mathrsfs}
\usepackage{tikz}
\usepackage{graphicx}
\usepackage{subfigure}
\usepackage{overpic}
\usepackage{pgfplots}
\title{Triangulations for ABHY Polytopes and Recursions for Tree and Loop Amplitudes}
\begin{document}

\author[a,b]{Qinglin Yang}
\affiliation[a]{CAS Key Laboratory of Theoretical Physics, Institute of Theoretical Physics, Chinese Academy of Sciences, Beijing 100190, China}
\affiliation[b]{School of Physical Sciences, University of Chinese Academy of Sciences, No.19A Yuquan Road, Beijing 100049, China}
\emailAdd{yangqinglin@itp.ac.cn}
\abstract
{In this note we make a field-theoretical derivation of a series of new recursion relations by a one-parameter deformation of kinematic variables for tree and one-loop amplitudes of bi-adjoint $\phi^3$ theory. Tree amplitudes are given by canonical forms/functions of associahedra realized in kinematic space by Arkani-Hamed, Bai, He and Yan (ABHY); the construction has been extended to generalized associahedra, where type $\mathcal{B}$/$\mathcal{C}$ polytopes compute tadpole diagrams and type $\mathcal{D}$ polytopes compute one-loop planar $\phi^3$ amplitudes. The new recursions are natural generalizations of the formula we found in \cite{He:2018svj}, and are shown to work for all ``C-independent" ABHY polytopes. Geometrically, the formula indicates triangulation of the generalized associahedron by projecting the whole polytope onto its boundary determined by the deformation. When projecting onto one facet, our recursion gives {\it e.g.} ``soft-limit triangulation"  and  ``forward-limit triangulation"  for tree and one-loop level. But we also find a lot of new formulae from our recursion relation, by projecting onto lower dimensional facets.}

\maketitle
\section{Introduction}
Recent years, certain physical theories and their amplitudes are found to be in deep connection with a new mathematical subject called {\it positive geometry} and its {\it canonical form/function} \cite{Arkani-Hamed:2017tmz}. In this picture, amplitudes of $\mathcal N=4$ SYM theory can be computed from the canonical forms of certain positive geometries which are called {\it amplituhedra} \cite{Arkani-Hamed:2013jha,Arkani-Hamed:2017vfh}, and each term of the famous BCFW recursion formula \cite{Britto:2004ap,Britto:2005fq} is related to a cell in ``triangulation" of such a geometrical object \cite{Hodges:2009hk,ArkaniHamed:2010gg}, turning the computation of $\mathcal N=4$ SYM theory amplitudes into a purely mathematical problem. In \cite{Arkani-Hamed:2017mur}, it was discovered that n-point planar tree-level bi-adjoint $\phi^3$ theory \cite{Cachazo:2013iea} can also be tied to a geometrical object called {\it associahedron} (denoted as $\mathcal{A}_{n-3}$) \cite{Stasheff_1,Stasheff_2}. Compared with amplituhedra for $\mathcal N=4$ SYM theory, associahedra have much simpler structure: they are polytopes living in kinematic space, defined in positive region of planar variables $X_{ij}=s_{i,i+1,\cdots,j-1}\geq0$ with linear constraints called ABHY conditions \cite{Arkani-Hamed:2017mur}. Each facet of such a polytope corresponds to a physical pole of the amplitude, and each vertex an n-point cubic Feymann diagram. In this geometrical viewpoint, tree level planar $\phi^3$ scattering forms, which arise naturally from worldsheet by scattering equation \cite{Cachazo:2013gna,Cachazo:2013hca}, turn out to be the canonical forms of the new mathematical objects. Furthermore, ABHY associahedron  also tightly links to several physical and mathematical topics and gives us many novel ideas over them, for instance moduli space of string worldsheet and type ${A}$ cluster and so on \cite{Arkani-Hamed:2017mur,He:2018pue}.

Most recently, idea of ABHY formalism has been extended to a wider range of combinatorial objects with physical background, for example all finite type cluster polytopes \cite{Bazier}, and made success in finding amplituhedron for one-loop level bi-adjoint $\phi^3$ theory \cite{Arkani-Hamed:2019vfh}. Denoted as $\bar{\mathcal{D}}_n$, its canonical function gives the integrand of one-loop $\phi^3$ amplitude. Combinatorially, such an object is interpreted as ``half of type $\mathcal{D}_n$ cluster polytope". Similar to construction of  type $\mathcal{A}$ associahedron, $\bar{\mathcal{D}}_n$ lives in kinematic space as well and its ABHY realization can be read out directly from mesh diagram \cite{Arkani-Hamed:2019vfh}. Furthermore, other kinds of finite type associahedra, {\it i.e.} type $\mathcal{B}$/$\mathcal{C}$ cluster polytopes (known as cyclohedra), are also found to be the amplituhedron for all tadpole diagrams after a similar ABHY realization \cite{Arkani-Hamed:2019vfh}, which is denoted as $\mathcal{B}_{n-1}$ in this note. Together with ABHY polytopes $\mathcal{A}$ and $\bar{\mathcal{D}}$, they are usually called ``generalized ABHY associahedra", which are recently related to topics like generalized string amplitudes and their moduli spaces \cite{Li:2018mnq,Arkani-Hamed:2019jha}, stringy integrals \cite{Arkani-Hamed:2019cfb}, {\it etc.}.

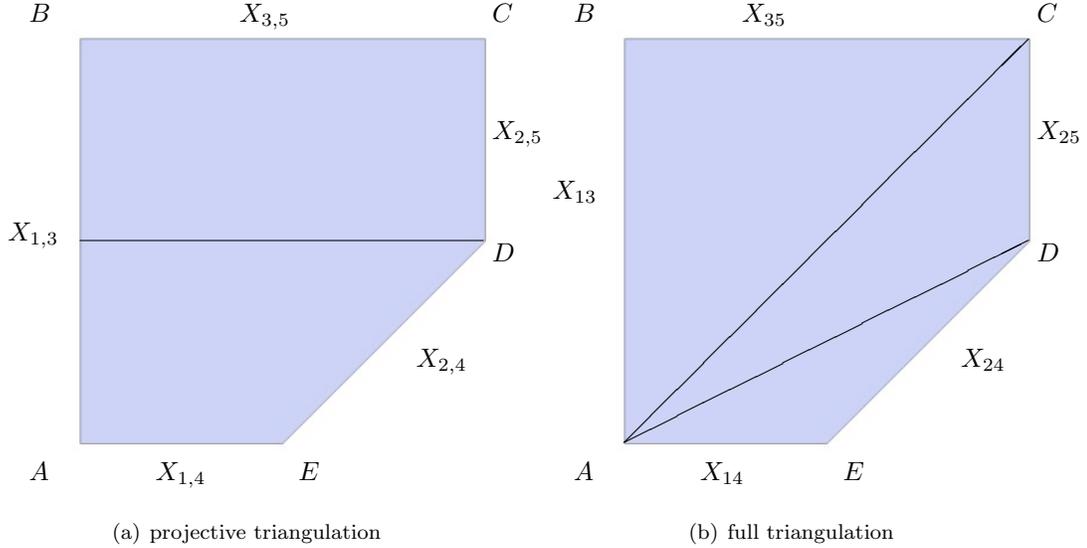
\begin{figure}
\centering
\subfigure[projective triangulation]{
\begin{minipage}{0.3\linewidth}
\centering
\begin{overpic}[width=1.2\textwidth]
{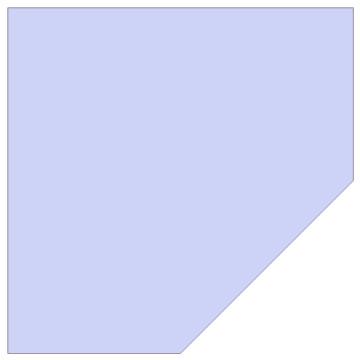}
\put(20,-7){$X_{1,4}$}
\put(-15,50){$X_{1,3}$}
\put(40,102){$X_{3,5}$}
\put(100,74){$X_{2,5}$}
\put(82,20){$X_{2,4}$}
\put(-10,-7){$A$}
\put(-10,102){$B$}
\put(100,102){$C$}
\put(100,45){$D$}
\put(54,-7){$E$}
\put(2,50){\color{black}\line(1,0){96}}
\end{overpic}
\linebreak
\linebreak
\end{minipage}
}
\phantom{aaaaaaaaaaaa}
\subfigure[full triangulation]{
\begin{minipage}{0.3\linewidth}
\centering
\begin{overpic}[width=1.2\textwidth]
{5pt.jpg}
\put(20,-7){$X_{14}$}
\put(-15,60){$X_{13}$}
\put(30,102){$X_{35}$}
\put(100,74){$X_{25}$}
\put(82,20){$X_{24}$}
\put(-10,-7){$A$}
\put(-10,102){$B$}
\put(100,102){$C$}
\put(100,45){$D$}
\put(54,-7){$E$}
\put(2,2){\color{black}\line(1,1){96}}
\put(2,2){\color{black}\line(2,1){96}}
\end{overpic}
\linebreak
\linebreak
\end{minipage}
}
\caption{Two kinds of studied geometrical triangulations for associahedron}\label{figure1}
\end{figure}

The underlying $\phi^3$ amplituhedron is also a powerful tool in computing amplitude of $\phi^3$ theory. According to property in positive geometry theory, we can always divide the whole polytope into several pieces, sum of whose canonical functions giving the final answer as canonical functions are geometrically additive. It is first studied in \cite{Arkani-Hamed:2017mur} for associahedron $\mathcal{A}_{n-3}$ geometrically that one can triangulate the associahedron by connecting one vertex with all other vertices and do the canonical function/tree level $\phi^3$ amplitude computation. As facets of associahedra $\mathcal{A}_{n-3}$ are always products of lower points ones, such a triangulation can be done recursively until the whole associahedron is divided into sum of simplices (called the full triangulation). In \cite{He:2018svj} we found a ``BCFW"-like recursive formula for full triangulation. After recursion of $n-3$ steps, tree level $\phi^3$ amplitude was represented as a sum of canonical functions of simplices, in analogy with the all-multiplicity solution of the BCFW recursion for $\mathcal{N}=4$ SYM theory \cite{Drummond:2008cr}. Besides dividing the associahedron into simplices, other kinds of triangulations were also discovered recently. In \cite{Arkani-Hamed:2019vfh}, a kind of new triangulations for generalized associahedra was introduced, called ``soft-limit triangulation" for tree level and ``forward-limit triangulation" for loop-level. Derived from a projection of amplituhedron onto a particular facet, it divides the polytope into several prisms, whose canonical functions are computed by canonical functions of their bottoms multiplying the heights. The procedures can be done recursively likewise, inducing a new geometrical recursion which is much more efficient than the recursion from triangulating the polytope into simplices (figure \ref{figure1}). These triangulations and their generalization, {\it i.e.} triangulations derived from projections, will be called ``{\it projective triangulation}" in this note. 

A natural question therefore arises that can we make a field-theoretical derivation of this recursive formula like the one we found in \cite{He:2018svj}, whose geometrical interpretation is projective triangulation? Moreover, due to the fact that the triangulation from one vertex is equivalent to projecting the polytope onto that vertex and dividing the associahedron, if we can find any relations between these two kinds of recursion formulae? Besides, are there any new recursive expressions for $\phi^3$ amplitudes and canonical functions of generalized ABHY associahedra? This note is written to give the answers. We will derive a series of new formulae for generalized ABHY associahedra (or more generally, for ``C-independent" ABHY polytopes) which are natural generalizations of our old recursion relation in \cite{He:2018svj}. Similar to the recursion before, the new relations are based on deformation $\{X_{A_i}\to zX_{A_i}\}_{i=1,\cdots,k}$ for arbitrary many (no more than dimension of the polytope) planar variables this time. An analog of the function we consider in \cite{He:2018svj}, which is also constructed from the canonical function of the polytope, will be studied. It will be proven that this function, as meromorphic function of the parameter $z$, never have pole at $z=\infty$. As a result, the amplitude/integrand, as residue at $z=1$ of the function, can always be represented as a sum of residues at finite poles $z_i$s by Cauchy theorem, which can be computed from canonical functions of the facets. Finally, the computation leads to new recursion relations of amplitudes due to factorization property of amplitudes. Especially in the two special cases that $k=$(dimension of the polytope) (also the maximally many variables we can rescale) and $k=1$, we will see that the formula will be coincident with the two recursions relation derived geometrically: triangulating the polytope by simplices and by prisms. Also, the first case, when only considering the tree level amplituhedra $\mathcal{A}_{n-3}$, is identical to the formula we derived in \cite{He:2018svj}.

Furthermore, the geometrical meaning of these recursions will be discussed in this note. We will give a proof, based on a formula presented in appendix of \cite{Arkani-Hamed:2019vfh}, and examples in tree level to illustrate the fact that when $k=1$ our new formula indeed indicates a projective triangulation for generalized associahedra. These formulae are the least complex ones throughout all the recursions, as they introduce least many residue terms. Moreover, situation of more than one (and less than dimension of the amplituhedron many) variables rescaled will also be discussed. Based on such a deformation, our derivation leads to a series of new recursion relations that are previously unknown, whose geometrical meaning is to project the polytope onto the lower dimensional facet determined by  deformed variables. As a triangulation, we will see that these new recursions have amazing structure: it divides the whole polytope into positive geometries with {\it curvy }facets. Finally, we will also discuss cases of loop level. One-loop amplituhedron $\bar{\mathcal{D}}_n$ for $\phi^3$ theory, comparing to tree level $\mathcal{A}_{n-3}$, have some significant features and differences, which will introduce new complexity for recursive computation. We also give more examples of one-loop integrand rewritten by different recursion relations. We will see that it is natural to derive the particular expressions from this formalism that loop integrands can be represented by purely tree amplitudes, which are called forward limits studied in \cite{Arkani-Hamed:2019vfh} from geometrical point of view. Also we can compute the integrands from tree amplitudes and sum of tadpole diagrams, after a different recursion based on new deformation and basis chosen.

The note is organized as following: the derivation of the formulae and some discussions over it are presented in section $2$, containing the proof of the deformation introducing no pole at infinity. In section $3$, we will discuss geometrical meaning of our formulae. Some specific examples will be offered to illustrate that the formulae indeed leads to triangulations from projections. Finally, in section $4$ we will extend our discussion to loop level.  We end the note with some discussions of the unsolved topics and some explicit results of computation examples mentioned throughout the note.

\subsection{Review and notations}

Let's firstly give a brief review of the tree level and $1$-loop amplituhedron of $\phi^3$ theory, {\it i.e.}, generalized associahedra $\mathcal{A}_{n-3}$ (for $n$-point) and $\bar{\mathcal{D}}_n$ (for $n$-point) \cite{Arkani-Hamed:2019vfh}. Throughout this note, we will describe these two polytopes purely from ABHY formalism, avoiding any cluster algebraic languages. As ABHY polytopes, they are both positive geometries defined by the planar variables $X_A$ in kinematic space. We will use $\Omega_n^{arb}$ to denote the canonical form of an arbitrary $n$-point amplituhedron (may be of tree or $1$-loop level), and correspondingly $A_n^{arb}$ for canonical function.

Firstly let's recall $\mathcal A_{n-3}$. We need a positive region:
\begin{equation}
\Delta_n=\{X_{i,j}\geq0\ {\rm for\ all}\ 1\leq i<j{-}1< n\}
\end{equation}
which is an $n(n-3)/2$-dim space. Here $X_{i,j}$ are defined by the extended Mandelstam variables $X_{i,j}=s_{i,i+1,\cdots,j-1}:=(p_i+\cdots+p_{j-1})^2$. $\mathcal A_{n-3}$ is then the intersection of this region $\Delta_n$ with $(n{-}2)(n{-}3)/2$ hyperplanes.
\begin{equation}\label{conditions}
\begin{split}
H(1,2,\cdots,n):=\{C_{i,j}=X_{i,j}+X_{i+1,j+1}-X_{i,j+1}-X_{i+1,j}\\
{\rm are\ positive\ constants, \, for}\ 1\leq i<j{-}1<n{-}1\}
\end{split}
\end{equation}
Therefore, at most $n-3$ of the planar variables are linearly independent and we can choose a proper basis (usually denoted as $\{X_{i_a,j_a}\}_{ a=1,\cdots,n-3}$ below) to represent all the $X_{i,j}$ by solving the conditions \eqref{conditions}. Especially, when we choose basis $\{X_{1,i}\}_{i=3,\cdots,n}$, a general solution of the planar variables can be found in the appendix of \cite{Arkani-Hamed:2019vfh}:
\begin{equation}\label{solA}
X_{i,j}=\sum_{\substack{a=1\\b=i+1}}^{\substack{a=i-1\\b=j-1}}C_{a,b}+X_{1,j}-X_{1,i+1}
\end{equation}
As an ABHY polytope, its canonical function can be directly written down from the vertices \cite{Arkani-Hamed:2017mur}. Throughout the note, we will denote the canonical function,{\it i.e.} $n-$point tree level amplitude as $A_n^{tree}(X,C)$ or $A^{tree}_{1,2,\cdots,n}$, where ``$X$" stands for the basis variables and ``$C$" those positive constants.

As mentioned in introduction, a natural way to compute function $A^{tree}_{1,2,\cdots,n}$ is triangulating the associahedron and adding canonical function of each part together. In \cite{He:2018svj} we discovered a ``BCFW -like" recursion relation, which is in fact a full triangulation of the associahedron in geometrical point of view:
\begin{equation}\label{2}
{A_{1,2, \cdots ,n}(X, C)=\sum_{(a,b)\neq (i,j)'s} \frac{z_{a b}^{n-3}}{X_{a b}}A_{a, \cdots ,b{-}1,I}(z_{a b} X, C) \times A_{I, b, \cdots, a{-}1}(z_{a b} X, C)}
\end{equation}
It was derived from a rescaling $X_{i_a,j_a}\to z X_{i_a,j_a}$ for all the basis variables. In this note we will have a more general discussion over such kind of recursions, applying these recursion relations both to tree and $1-$loop level cases. 

To recall the $\bar{\mathcal{D}}_n$, we should review generalized ABHY associahedron $\mathcal{D}_n$ at first. In  ABHY formalism, all the $n^2$ facets of $\mathcal{D}_n$ can be sorted as: $\frac{n(n-3)}2$ $X_{i,j}$ and $\frac{n(n-3)}2$ $X_{j,i}$, where $1\leq i<j-1\leq n-1$. $n$ $X_{i,i+1}$, and cut facets $n$ $X_i$ and $n$ $X_i^\prime$. The polytope is defined as the intersection of the positive region $\{X\geq0\}$ with the $n(n-1)$ constraints \cite{Arkani-Hamed:2019vfh}:
\begin{equation}\label{Dbarn}
\begin{split}
X_{i,j}+X_{i+1,j+1}-X_{i,j+1}-X_{i+1,j}=C_{i,j}\ \ \ for\ |j-i|>1, i\neq n\\
X_{i,i+1}+X_{i+1,i+2}-X_{i,i+2}-X_{i+1}-X_{i+1}^\prime=C_{i,i+1}\ \ \ for\ 1\leq i<n\\
X_i+X_{i+1}^\prime-X_{i,i+1}=C_i,\ \ \ X_i^\prime+X_{i+1}-X_{i,i+1}=C_i^\prime\ \ for\ 2\leq i\leq n
\end{split}
\end{equation}
which then becomes an $n-$dim polytope. Combinatorially, however, the number of the vertices is not equal to the number of $1-$loop Feymann diagrams: all the cut facets are doubled, so we need to subtract half of them \cite{Arkani-Hamed:2019vfh}. Introduce two new variables $X_+=X_1^\prime+X_1$ and $X_-=X_1^\prime-X_1$ (in fact we also have $X_-=X_i^\prime-X_i$ by this definition). The facet $X_-$ then divides the $\mathcal{D}_n$ polytope into two identical parts, each of which is what we called the generalized associahedron $\bar{\mathcal{D}}_n$ (facets $\{X_-,X_i,X_{i,i+2},X_{i,j},X_{j,i}\}$). This polytope, whose vertices one-to-one correspond to the $1-$loop diagrams, is then regarded as the $1-$loop amplituhedron of the $\phi^3$ theory.

Following from the conditions \eqref{Dbarn} and the definition of $X_{\pm}$, we can then represent the facets of $\bar{\mathcal{D}}_n$ polytope by the basis $\{X_+,X_-,X_{1,i}\}_{i=2,3,\cdots,n-1}$,which is called the tadpole basis (appendix of \cite{Arkani-Hamed:2019vfh}):
\begin{equation}\label{solD}
\begin{split}
&X_i=c_A+X_{1,i}-\frac12(X_++X_-)\\
&X_{i,j}=2c_A+c_B+X_{1,i}+X_{1,j}-X_+\\
&X_{j,i}=c_D+c_B+2c_C+X_{1,j}-X_{1,i+1}
\end{split}
\end{equation}
where $c_{A,B,C,D}$ are sums of constants $C_{ij}$.

Finally, for readers easier to refer, we here briefly list the facet structure of $\bar{\mathcal{D}}_n$. Physically they stand for the factorization properties of the integrand of $1$-loop amplitude. The cut facets $X_i=0$ are associahedra $\mathcal{A}_{n+2}$, which physically mean forward limits of the loop amplitude:
\begin{equation}
\lim_{X_i\to 0} X_i A_n^{1-loop}(1,2,\cdots,n)=A_{n+2}^{forward\ tree}(1,2,\cdots,i,-,+,i+1,\cdots,n)
\end{equation}
When $X_{i,i+1}\to 0$, the polytope factorizes into a direct product $\mathcal{A}_n\times\bar{\mathcal{D}}_2$. For arbitrary $X_{i,j}\to0\ (i<j-1)$ (or $X_{j,i}\to0\ (j>i+1)$), polytope factorizes as $\mathcal{A}_{i,i+1,\cdots,j}\times\bar{\mathcal{D}}_{1,2,\cdots,i,j,\cdots,n}$ (or $\mathcal{A}_{1,2,\cdots,i,j,\cdots,n}\times\bar{\mathcal{D}}_{i,i+1,\cdots,j}$). Lastly, facet $X_-=0$ is a $\mathcal{B}_{n-1}$ polytope, which means that we have a tadpole limit:
\begin{equation}
\lim_{X_-\to 0}X_-A_n^{1-loop}(1,2,\cdots,n)=A_n^{tadpole}(1,2,\cdots,n)
\end{equation}
where $A_n^{tadpole}(1,2,\cdots,n)$ denotes the sum of all one-loop level tadpole diagrams. All these properties play a crucial role in our new recursions below.

\section{Recursion relation for $\phi^3$ amplitude from projection}
\subsection{Derivation of the recursions}
As a natural generalization of the recursion \eqref{2}, let's consider a general rescaling 
\begin{equation}\label{1}
X_{A_i}\to z X_{A_i} \ \ \ i=1...k
\end{equation}
for $n-$point arbitrary amplitude/integrand (briefly amplitude below) $A_n^{arb}(zX,C)$ which comes to be a meromorphic function of $z$. Here $X_{A_i}$ are some linearly independent planar variables, or facets forming a generalized ABHY associahedron, and $1\leq k\leq $(dimension of the polytope). Without loss of generality, we can just choose these variables as part of the basis $\{X_{A_i}\}_{i=1,\cdots,n}$. Note that we need not require whether these deformed variables are compatible. Similar to the recursion in \cite{He:2018svj}, in this case we consider the integral:
\begin{equation}
\oint \frac {z^k}{z-1}A_n^{arb}(zX,C)
\end{equation}
Original amplitude is given by the residue at $z=1$. By Cauchy theorem, the amplitude is written as:
\begin{equation}
A_n^{arb}(X, C)=-\left({\rm Res}_{\infty} + \sum_ {\rm finite\ poles} {\rm Res}_{z_i}\right)(\frac{z^k}{z-1}A_n^{arb}(zX,C))
\end{equation}
As we put a $z^k$ on the numerator, following from the discussion in \cite{He:2018svj} similarly, the function $\frac{z^k}{z-1}A_n^{arb}(zX,C)$ won't have pole at $0$.

It seems that the behavior of the function at infinity now is more complicated to illustrate after a more general deformation. Fortunately, such a function always behaves well even after an arbitrary rescaling. In fact we have a property: For rescaling of the variables:
\begin{equation}
X_{A_i}\to z X_{A_i} \ \ \ i=1...k
\end{equation}
here $\{X_{A_i}\}_{i=1,\cdots,n}$ can be arbitrary basis chosen when solving the ABHY conditions of generalized ABHY associahedra (or more generally, arbitrary polytope that can be constructed by ABHY formalism\footnote{Note that here is a nontrivial prerequisite: We always suppose that the ABHY polytope is ``C-independent", which means that the combinatorial shape of the polytope won't be affected by the variation of constants $C$. Therefore list of its vertices will also not be affected by values of constants $C$, and we can directly write down its canonical function from its vertices. }). Then function $\frac{z^k}{z-1}A_n(zX,C)$, where $A_n(X,C)$ is the canonical function of the polytope, has no pole at infinity. This property will be proved in the later part of this section. 

Armed with such a property, we can therefore only take the residue of the function at each finite pole $z_i$ into account. After all the other planar variables are represented by variables in chosen basis $\{X_{A_i}\}_{i=1,\cdots,n}$, these $z_i$ are solutions of equations $\hat X_{B_i}(z)=0$ for planar variable $X_{B_i}$s that linearly depend on rescaled basis variables $X_{A_i}$. We generally have a representation of $X_{B_i}$ as:
\begin{equation}\label{solarb}
X_{B_i}=c_{B_i}+X_i+\sum_{j=1}^k\lambda_{i,j} X_{A_j}
\end{equation}
where $\lambda_{i,j}$ are real numbers, $c_{B_i}$ and $X_i$ are linear combinations of constants $C$s and undeformed variables $\{X_{A_j}\}_{j=k+1,\cdots,n}$ in basis. Supposing at each physical pole the amplitude factorizes as:
\begin{equation}
\lim_{\hat X_{B_i}\to 0}\hat X_{B_i}(z)A_n(zX,C)=A^{limit}_{B_i}(z_{B_i}X,C)
\end{equation}
the residue then reads:
\begin{equation}
Res_{\hat X_{B_i}}\frac{z^k}{z-1}A_n^{arb}(zX,C)=\frac{z_{i}^{k}}{(\sum_{j=1}^k\lambda_{ij} X_{A_j})(z_i-1)}A^{limit}_{B_i}(z_{B_i}X,C)
\end{equation} 
Following the formula \eqref{solarb}, we finally arrive at the result:
\begin{equation}\label{recursion}
\boxed{ A_n(X,C)=\sum_{B_i}\frac {z_i^k}{X_{B_i}}A^{limit}_{B_i}(z_{B_i}X,C)}
\end{equation}
where $B_i$ runs over the planar variables shifted by the rescaling \eqref{1}. It can be seen that \eqref{recursion} is a generalization of the recursion \eqref{2} we found before. When considering tree level amplitude with $k=n-3$, \eqref{recursion} and \eqref{2} are coincident. (Note that for $n$-point tree amplitude, its amplituhedron $\mathcal{A}_{n-3}$ is an $n-3$ dimensional polytope. So that $k=n-3$ is also the maximal number of planar variables we can rescale in our formualism.)  At this point, for an $n$-dim generalized associahedron as amplituhedron  and $k=n$ {\it i.e.} deforming all the (maximally many) planar variables in the basis like what we did to derived \eqref{2}, the recursive relation \eqref{recursion} reads:
\begin{equation}\label{recursion3}
A_n(X,C)=\sum_{B_i}\frac {z_{B_i}^n}{X_{B_i}}A^{limit}_{B_i}(z_{B_i}X,C)
\end{equation}
Here $X_{B_i}$ runs over all the planar variables except the variables $X_{A_i}$ in basis. And formula \eqref{recursion3} can be viewed as a direct generalization of our old formula \eqref{2} by simply applying \eqref{2} to generalized associahedra. Differences among new formulae  and original \eqref{2} are that now in \eqref{recursion} and \eqref{recursion3} $A^{limit}$ can be a product of tree, loop level amplitudes/integrands or even sum of tadpole diagrams, depending on the property of the amplituhedron we take into account, and in \eqref{recursion} $k$ on the numerators of each term is variable, determined by the number of the planar variables we rescale.

Besides, there is also a particular and interesting case when we consider a one-variable rescaling:
\begin{equation}
X_A \to zX_A
\end{equation}
Here $k=1$, and we can go on simplifying \eqref{recursion}. Suppose $X_{B_i}\propto \lambda_iX_A$, where $\lambda_i$ is a real number. Then 
\begin{equation}
\frac {z_A}{X_{B_i}}=(\frac 1{X_{B_i}}-\frac1{\lambda_iX_A})
\end{equation}
giving the final result:
\begin{equation}\label{recursion2}
\boxed{A_n(X,C)=\sum_{B_i}(\frac 1{X_{B_i}}-\frac1{\lambda_iX_A})A^{limit}_{B_i}(z_{B_i}X,C)}
\end{equation}
In section $3$ we will see that this formula indicates a projective triangulation of generalized associahedron. 

Last thing we should mention in this section is that physical poles $X_{B_i}$ we need to consider in a computation, even when rescaled $X_{A_i}$s are determined, {\it will} depend on the other undeformed variables in basis $\{X_{A_i}\}_{i=k+1,\cdots,n}$ we choose. A simple example is the $1-$loop amplituhedron: When choosing tadpole variables $\{X_+,X_-,X_{1,j}\}$, according to the result \eqref{solD}, variables $X_{i,j}$ don't depend on the loop variables $X_i$. However, if we change the basis to $\{X_1,X_-,X_{1,j}\}$ they do. It can be comprehended easily in a geometrical point of view. After a specific group of basis variables are chosen, we in fact suppose that they are vertical to each other geometrically. Angles formed by facets therefore depend on such a selection of basis, so do the image of the facets after projection. We should thus firstly make it clear that which group of basis variables have we chosen before using \eqref{recursion} to compute amplitude. In the absence of special instruction, we will always choose $\{X_{1,i}\}_{i=3,\cdots,n}$ in tree level case and tadpole variables $\{X_+,X_-,X_{1,j}\}_{j=1,\cdots,n-1}$ in $1-$loop case as basis variables to represent the others.

\subsection{Proof of no pole at infinity}
In the end of this section, we present the postponed proof of the property: {\it For arbitrary polytope constructed from ABHY conditions, deformation \eqref{1} introduces no pole at infinity for function $\frac{z^k}{z-1}A_n^{arb}(zX,C)$.} So that our formulae can be applied to arbitrary ABHY polytopes\footnote{However, as general ABHY polytopes may not satisfy the property the generalized associahedra have {\it i.e.} facets of generalized associahedra are products of lower points ones, the formulae won't turn out to be  recursion relations in general cases. Merely in this subsection will we extend our discussion beyond generalized associahedra.}. This proof is divided into two steps: In the first step we prove a condition called soft condition as a lemma, which is a crucial property with which will the residue at infinity vanishes. Then in the second part, the main property is shown. As has been noted in the footnote above, whenever we mention ``arbitrary ABHY polytope" in this section, we always assume it is ``$C$-independent".
\paragraph{Proof of the soft condition}
To begin with, we should illustrate the meaning of the soft condition. An arbitrary ABHY polytope always satisfy the condition:
\begin{equation}\label{soft}
\lim_{C\to 0}A_k(X,C)=0
\end{equation}
To prove this condition, suppose we are now taking into account an ABHY polytope defined as the positive region $\{X_{A_i}\geq0\}_{i=1,\cdots,n}$ constrained by $k$ ABHY conditions:
\begin{equation}\label{ABHY}
f_a(X_{A_i})=C_a\ \ \ for\ a=A_{i_1},\cdots,A_{i_k} 
\end{equation}
where $f_a(X)$ are $k$ linear functions of these planar variables. Note that we need only consider the situation that these conditions are linearly independent, or we can just delete the redundant ones to make it back to the independent case. These ABHY conditions together with the positive region then form an $n-k$-dim polytope. Solving these conditions we can represent every planar variables by the variables as basis, without loss of generality, $\{X_{A_i}\}_{i=1,\cdots,n-k}$:
\begin{equation}
X_{A_i}=g_{A_i}(X_{j},C_{A_l})
\end{equation}
for $i=n-k+1,\cdots,n,\ \ \ j=1,2,\cdots,n-k$ and $l=1,\cdots,k$. After taking the limit $C_{A_j}\to 0$, $g_{A_i}$ become only linear functions of variables $\{X_{A_i}\}_{i=1,\cdots,n-k}$.

Now think of the coordinate of arbitrary vertex $X_{A_{i_1}}X_{A_{i_2}}\cdots X_{A_{i_n-k}}$ under such a solution and the limit $C_{A_i}\to0$. We should solve the linear equations:
\begin{equation}\label{vertexeq}
g_{A_i}(X_{A_j},0)=0\ \ \ \ for\ i=i_1,\cdots,i_{n-k}
\end{equation}
If the condition \eqref{soft} is not satisfied by this ABHY polytope, as a result, there must be at least a group of linear equations \eqref{vertexeq} has non-zero solutions. It contradicts the fact that, to form a vertex, every group of linear equations \eqref{vertexeq} must also be of full rank. 

We can also illustrate limit \eqref{soft} in geometrical language. In a geometrical viewpoint, those constants $C_{A_i}$ control the ``size" of the ABHY polytope. So when the limit $C_{A_i}\to0$ is taken, geometrically it means that the polytope shrinks to the origin, whose canonical function will of course becomes $0$.

\paragraph{Proof of no pole at infinity}
Now we turn to the main property. Suppose we rescale basis variables
\begin{equation}
X_{A_i}\to z X_{A_i} \ \ \ i=1...k
\end{equation}
for an arbitrary ABHY simple polytope $\mathscr{A}$, whose canonical function is $A_n(X,C)$. We know that $A_n$ must have the form
\begin{equation}
A_n(X,C)=\sum_{vertices}\frac1{\prod^n X}
\end{equation}
Here $\prod^n X$ in each term are those facets that share the vertex (Without loss of generality, suppose there are n $X$s in one denominator). To check the behavior of function $\frac {z^k}{z-1}A_n(zX,C)$ at infinity, we need only consider those terms that are of $O(\frac1{z^k})$ when $z\to\infty$, {\it i.e.}, $n-k$ $X$s (denoted as $X_{{v_j}}$ below) in their denominators are vertical to all those basis we rescale. They are unshifted after the deformation. As these $X_{v_j}$ should also be linearly independent to each other, there won't be any terms of $O(\frac1{z^{k-1}})$ or less. (As $\mathscr{A}$ is an $n-$dim polytope, there cannot be more than $n-k$ planar variables that are linearly independent meanwhile vertical to deformed variables.)

Now for the terms considered, we can group them by these $X_{v_j}$: terms having the same $n-k$ $X_{v_j}$s will be putted into the same group. Then each group of terms adds up to make a contribution:
\begin{equation}
(\sum_{\substack{vertices\ shared\\\ by\ n-k\ X_{v_j}s}}\frac1{\prod^{k}X})\frac1{\prod_j^{n-k}X_{v_j}}
\end{equation}
to the whole canonical function, and only the part in the brackets is affected by the rescaling. In fact, sum in the bracket above is just the canonical function of the factorizd ABHY polytope when all these $X_{v_j}\to 0$, {\it i.e.} the canonical function of the intersection of all these facets $X_{v_j}$, which we denote as $A_k(X,C)$ here. Thus when $z\to\infty$, $\lim_{z\to\infty} A_k(X,C)\sim \frac1{z^{k}}A(X,0)$, and this part will go to $0$ by the soft limit: 
\begin{equation*}
\lim_{C\to 0}A_k(X,C)=0
\end{equation*}
This property is held for every ABHY polytope, which is shown in the first part of the proof. Thus $A_k(X,C)\sim O(\frac1{z^{k+1}})$ when $z\to\infty$. The whole residue of function $\frac{z^k}{z-1}A_n(zX,C)$ at infinity thus vanishes, which was to shown.

After finishing the proof, now let`s turn to some explicit examples to illustrate the recursion.

\section{Recursions in a geometrical viewpoint}
In this section, we will turn to the geometrical meanings of the relations \eqref{recursion} and \eqref{recursion2}. In first part of this section, we will show the geometrical interpretation of formula \eqref{recursion2}, which can be proved generally. We will also use an explicit case, the tree level amplituhedron $\mathcal{A}_{n-3}$, to illustrate such a property. In the later part of this section, deformation \eqref{1} lying in the middle case {\it i.e.} $1<k<$(dimension of the polytope) will also be taken into consideration, which will be illustrated by a specific example: $n=6$ tree level amplitude and its amplituhedron $\mathcal{A}_3$. As we have mentioned, these recursions \eqref{recursion} are totally new and will give us many unknown recursion relations for tree amplitudes and $1-$loop level integrands. Geometrically we will see that these formulae have a quite similar geometrical meaning as \eqref{recursion2}: triangulating the amplituhedra by projection onto its boundary determined by intersection of all deformed variables.

Before turning to the new ones, we should give a discussion over geometrical interpretation for expression \eqref{recursion3}:
\begin{equation*}
A_n(X,C)=\sum_{B_i}\frac {z_{B_i}^n}{X_{B_i}}A^{limit}_{B_i}(z_{B_i}X,C)
\end{equation*} 
very quickly. As we have mentioned in section 2.1, this formula is a generalization of the old recursion relation \eqref{2} if we consider generalized associahedron of $n$-dimension  based on also maximal many variables rescaled. Geometrically speaking, It indicates a triangluation of generalized associahedra from one vertex, as we discussed in detail in our previous paper \cite{He:2018svj} for tree level case, and the discussion there can be extended to generalized associahedra without difficulties. After n-step of recursive computation, we can solve the recursion relation and express amplitude/integrand into a sum of ``R-invariant" like functions, which are geometrically canonical functions of simplices produced by a full triangulation of the amplituhedra.

However, an apparent observation is that the fewer variables we rescale, the fewer physical poles come out, and thus fewer residues we need to compute during each step of the recursion. Therefore, it can be inferred that \eqref{recursion3} is the most complex formula throughout all kinds of recursions \eqref{recursion} as we rescale maximally many variables. The formula \eqref{recursion2}, on the other hand, should be the simplest. Now let's turn to discussions over it. 

\subsection{Recursion from one variable rescaled}\label{3.1}
Although the geometrical interpretation of \eqref{recursion} is still unclear, we are now able to illustrate \eqref{recursion2} in geometrical language. In fact, doing a recursive computation \eqref{recursion2} is equivalent to the ``projective triangulation": Firstly project the whole ABHY polytope onto the hyperplane determined by function $X_{A}=0$. divide the polytope into several pieces of prisms formed by shifted variables and their image as the top and the bottom. Terms on the RHS of \eqref{recursion2} are just the canonical functions of these prisms. After a direct addition, we get the final result $A_n^{arb}$.

To prove this property, it should at first be mentioned that, evaluating function $A_n(zX,C)$ at each $z_i$ determined by the solution of $X_{B_i}(z_i)=0$ is equivalent to make a substitution in original factorized amplitude $A^{limit}_{B_i}(X,C)$ as:
\begin{equation}\label{change}
X_{A}\to X_{A}-\frac1{\lambda_i}X_{B_i}
\end{equation}
for each $X_{B_i}\propto \lambda_iX_A$ term, which can be checked directly by computation. In appendix of \cite{Arkani-Hamed:2019vfh}, a formula for the canonical function of prism $\mathcal{P}$ with $\mathcal{U}$pper and $\mathcal{B}$ottom facets has been presented as:
\begin{equation}\label{prism}
A_{\mathcal{P}}(Y)=(\frac{Z\cdot W_{\mathcal{B}}}{Y\cdot W_{\mathcal{B}}}-\frac{Z\cdot W_{\mathcal{U}}}{Y\cdot W_{\mathcal{U}}})A_{\mathcal{U}}(Y-\frac{Y\cdot W_{\mathcal{U}}}{Z\cdot W_{\mathcal{U}}}Z)
\end{equation}
where $Z$ is an arbitrary reference point, $Y=(1,X_{1,3},\cdots,X_{1,n})$ in our situation and $W$s are dual vectors of $\mathcal{P}$`s upper and bottom facets. Comparing to this result, replacement \eqref{change} in fact leads to a choice of reference point $Z=(0,\frac1{\lambda_i},0,\cdots,0)$. After identifying $Y\cdot W_{\mathcal{U}}$ and $Y\cdot W_{\mathcal{B}}$ in each term with $X_{B_i}$ and $X_{A}$ respectively and applying $Z$, one can figure out absolutely same results from terms in \eqref{recursion2} and equation \eqref{prism}, which proves the property. In the following, when considering a one-variable-rescaling, we will always use the language: doing a substitution \eqref{change} in each factorized amplitude $A^{limit}(X,C)$, instead of evaluating each term of \eqref{recursion2} at $z_a$, to avoid introducing an extra variable $z$.

Now, as a concrete example, let`s consider a one-variable-rescaling for $\mathcal A_{n-3}$: 
\begin{equation}
X_{1,3}\to z X_{1,3}
\end{equation}
By the relations \eqref{solA}, only variables $X_{2,i}\propto - X_{1,3}$ will be shifted. Following the discussion in the last section, we can directly write down the recursion relation as:
\begin{equation}\label{tree1shift13}
A_{1,2,\cdots,n}^{tree}=\sum_{a=4}^n(\frac1{X_{1,3}}+\frac1{X_{2,a}})\hat{A}_{2,\cdots, a-1,I}^{tree}\times \hat{A}_{1,I,a,\cdots,n}^{tree}
\end{equation}
Here $A^{limit}$ in \eqref{recursion2} reads a factorization of tree amplitude. Hats on factorized amplitudes mean that before sum over $a$, we need to make the replacement \eqref{change}. In this case, we should substitute all the $X_{2,i}$ in the factorized amplitudes to $X_{2,i}-X_{2,a}$ in the $a$th term. This result is identical to the ``soft-limit triangulation" formula of the associahedron, which was discussed in \cite{Arkani-Hamed:2019vfh}.  As a projective triangulation, it has a specific feature that it is always an ``inside triangulation", {\it i.e.} all the prisms formed by the projection are inside the associahedron $\mathcal{A}_{n-3}$. This property can be read from ABHY realization of polytope $\mathcal{A}$: All the planar variables compatible to $X_{1,3}$ are those variables without index $2$, which are also linearly independent of $X_{1,3}$ according to the solutions \eqref{solA}. As a result, all the planar variables compatible to $X_{1,3}$ are also vertical to it, making the shadows of facets $X_{2,a}$ after projection always inside $X_{1,3}$, therefore forming an ``inside triangulation" of associahedron. Physically, as also discussed in \cite{Arkani-Hamed:2019vfh} geometrically, this rescaling is similar to the ``soft" limit: All the unshifted planar variables are those $X_{ij}$s whose indices $i,j\neq2$, which is like to ``forget" the vertex $2$ of the $n-$gon the associahedron $\mathcal{A}_n$ constructed from. These variables can then be viewed as planar variables of facet $X_{1,3}$ as factorized ABHY associahedron and form a associahedron of $n-1$ points. And we need only to consider the residues at poles apart from these, which are $X_{2,a}$s. So that, we also call such a limit ``forgotten limit".

Besides, equation \eqref{recursion2} now allows us to consider a more widely group of such projection recursions, {\it i.e.} projecting the associahedron onto arbitrary facet $X_{1,j}$. 
Generally, after $X_{1,j}$ being rescaled, variables:
\begin{equation*}
X_{a,j},\ \ a=2,\cdots,j-2 \ \ \ \& \ \ X_{j-1,a}\ \ a=j+1,\cdots,n
\end{equation*}
will be shifted. So in $\mathcal{A}$ cases, number of terms on the RHS of the sum \eqref{recursion2} is independent of which base we rescale, always $n-3$, and a similar computation gives us the result:
\begin{equation}\label{tree1shift}
\begin{split}
A_{1,2,\cdots,n}^{tree}=\sum_{a=2}^{j-2}(\frac1{X_{a,j}}-\frac1{X_{1,j}})\hat{A}_{a,\cdots, j-1,I}^{tree}\times \hat{A}_{1,\cdots a,I,j,\cdots,n}^{tree}
+\sum_{a=j+1}^n(\frac1{X_{j-1,a}}+\frac1{X_{1,j}})\hat{A}_{j-1,\cdots, a-1,I}^{tree}\times \hat{A}_{1,\cdots j,I,a,\cdots,n}^{tree}
\end{split}
\end{equation} 
Certainly every amplitudes with hat are under replacement \eqref{change} respectively.

\begin{figure}[htbp]
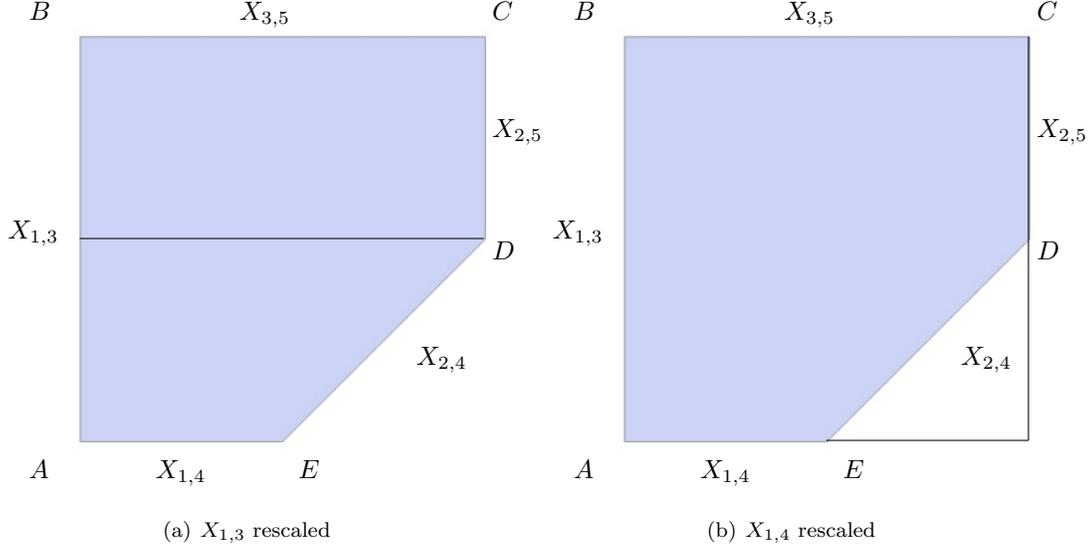

\centering
\subfigure[$X_{1,3}$ rescaled]{
\begin{minipage}{0.3\linewidth}
\centering
\begin{overpic}[width=1.2\textwidth]
{5pt.jpg}
\put(20,-7){$X_{1,4}$}
\put(-15,50){$X_{1,3}$}
\put(40,102){$X_{3,5}$}
\put(100,74){$X_{2,5}$}
\put(82,20){$X_{2,4}$}
\put(-10,-7){$A$}
\put(-10,102){$B$}
\put(100,102){$C$}
\put(100,45){$D$}
\put(54,-7){$E$}
\put(2,50){\color{black}\line(1,0){96}}
\end{overpic}
\linebreak
\linebreak
\end{minipage}
}
\phantom{aaaaaaaaaaaa}
\subfigure[$X_{1,4}$ rescaled]{
\begin{minipage}{0.3\linewidth}
\centering
\begin{overpic}[width=1.2\textwidth]
{5pt.jpg}
\put(20,-7){$X_{1,4}$}
\put(-15,50){$X_{1,3}$}
\put(40,102){$X_{3,5}$}
\put(100,74){$X_{2,5}$}
\put(82,20){$X_{2,4}$}
\put(-10,-7){$A$}
\put(-10,102){$B$}
\put(100,102){$C$}
\put(100,45){$D$}
\put(54,-7){$E$}
\put(98,2){\color{black}\line(0,1){96}}
\put(98,2){\color{black}\line(-1,0){48}}
\end{overpic}
\linebreak
\linebreak
\end{minipage}
}
\caption{Projective triangulations for $5$-point tree level amplituhedra with different variables rescaled}\label{5pt instri}
\end{figure}

As a simplest and explicit example, let`s briefly discuss a $5-$point situation. Firstly, consider the simplest case: $\{X_{13},X_{14}\}$ as basis with only $X_{13}$ rescaled. It means to project the pentagon onto the edge $X_{13}$. With the analysis above we can immediately write down the answer now:
\begin{equation}
A_5=(\frac1{X_{1,3}}+\frac1{X_{2,4}})(\frac1{X_{1,4}}+\frac1{X_{2,5}-X_{2,4}})+(\frac1{X_{1,3}}+\frac1{X_{2,5}})(\frac1{X_{3,5}}+\frac1{X_{2,4}-X_{2,5}})
\end{equation}
It is very easy for readers to check the answer is correct. The two terms stand for the canonical functions of the two $2-$dim prisms in the graph (figure \ref{5pt instri} (a)).

As for rescaling $X_{1,4}\to zX_{1,4}$, poles at $X_{2,4}$ and $X_{3,5}$ need to be considered now. We can also directly write down the answer:
\begin{equation}\label{4-pt2}
A_5=(\frac1{X_{3,5}}+\frac1{X_{1,4}})(\frac1{X_{1,3}}+\frac1{X_{2,5}})+(\frac1{X_{2,4}}-\frac1{X_{1,4}})(\frac1{X_{1,4}-X_{2,4}}+\frac1{X_{2,5}})
\end{equation}
The first term in the result, as a product of two canonical functions of segments, is the canonical function of the rectangle form by $X_{1,3}$ and $X_{3,5}$ as its edges. This is also the prism formed by projecting the facet $X_{3,5}$ onto line $X_{1,4}$. While the second term, appearing to be a $3-$term sum after the brackets expanded, is the canonical function of the triangle outside the associahedron with a minus sign. \eqref{4-pt2} then forms an ``outside" triangulation of the associahedron, which is the canonical function of a rectangle with a triangle {\it subtracted} (figure \ref{5pt instri} (b)). In all $n-3$ $X_{i_a,j_a}$ rescaled situation, such a triangulation will only appear when the basic point we choose is outside the associahedron, {\it i.e.} the rescaled $n-3$ $X_{i_a,j_a}$ are not compatible \cite{He:2018svj}.

\subsection{Recursion from arbitrary many variables rescaled}
After discussion over the equation \eqref{recursion2}, now let's turn to a more general case: to rescale arbitrary many $X_A$s and using the original formula \eqref{recursion} to compute the amplitude. We will also check the geometrical interpretation of such a recursion relation, by $\mathcal{A}_3$ with two variables rescaled as an example.

Now choosing basis $\{X_{1,3},X_{1,4},X_{1,5}\}$ and rescale $X_{1,3}$ and $X_{1,4}$ to $zX_{1,3}$ and $zX_{14}$, the sum then runs over $5$ terms: $X_{2,4}$, $X_{2,5}$, $X_{2,6}$, $X_{3,5}$ and $X_{3,6}$, which are those facets that are not both vertical to $X_{1,3}$ and $X_{1,4}$. The recursion \eqref{recursion} in this case reads:
\begin{equation}
A_{1,2,\cdots,6}^{tree}(X,C)=\sum_{i,j}\frac {z_{i,j}^2}{X_{i,j}}A_{i,i+1,...,j-1,I}^{tree}(z_{i,j}X,C)\times A_{1,\cdots,i-1,I,...,6}^{tree}(z_{i,j}X,C)
\end{equation}
where $z_i$ evaluates at solves of $\hat X_{i,j}(z)=0$. This is a specific example among a series of totally new representations for tree level $\phi^3$ amplitudes, which can be derived from the general formulae.

Let`s compute some terms in the sum as examples. Firstly we think of the term $X_{36}$:
\begin{equation}\label{36}
\Omega_{36}=\frac{z_{36}^2}{X_{36}}(\frac1{\hat X_{35}}+\frac1{\hat X_{46}})(\frac1{\hat X_{13}}+\frac1{\hat X_{26}})
\end{equation}
After a directly computation, this term reads:
\begin{equation}\label{36}
\Omega_{36}=N_{36}/(X_{13}X_{36}X_{46}(C_{15}+C_{25}-X_{15})(C_{24}X_{13} + C_{25} X_{13} + C_{14} (X_{13} - X_{14}) + C_{15} (X_{13} - X_{14}) - C_{13} X_{14}))
\end{equation}
where the $N_{36}$ is the numerator:
\begin{equation}
N_{36}=(C_{13}+C_{14}+C_{15})(C_{14}+C_{24}+C_{15}+C_{25})C_{35}
\end{equation}
This term has five linear poles in basis $X_{13},X_{14},X_{15}$, three of which are physical and two are spurious. It can be easily checked that the result \eqref{36} is in fact the canonical function of a triangular prism, which is formed by projecting the facet $X_{36}$ onto the line $X_{13}X_{14}$, {\it i.e.}, the positive geometry enveloping all the vertical lines from $X_{36}$ to $X_{13}X_{14}$. All its facets are planes in kinematic space and the spurious poles are then all linear functions in chosen basis, due to the fact that all the edges of facet $X_{36}$ are either vertical or parallel to line $X_{13}X_{14}$.

Next we consider $X_{26}$. With a totally same computation we obtain the result:
\begin{equation}\label{26}
\begin{split}
\Omega_{26}=N_{26}/(X_{26}X_{46}Y_1Y_2Y_3W)
\end{split}
\end{equation}
which is thus a canonical function of a positive geometry with six facets. First $5$ poles in \eqref{26} except $X_{26}$ itself ($X_{46}$, $Y_1$, $Y_2$ and $Y_3$, one can find the explicit results of these poles in the appendix) are also linear functions in basis $\{X_{13},X_{14},X_{15}\}$ and can be checked that they are planes formed by projection of four edges of facets $X_{26}$ that parallel or vertical to the target line (edge $X_{26}X_{25}$, $X_{26}X_{24}$, $X_{26}X_{46}$ and $X_{26}X_{36}$), like the situation $X_{36}$. 

However, there is still a spurious pole of twice power in basis:
\begin{equation}
W=-C_{14} X_{13} - C_{24} X_{13} + C_{13} X_{14} + C_{14} X_{14} + C_{15} X_{14} - X_{13} X_{15}
\end{equation}
and will be a {\it curvy surface} inside the associahedron! This spurious pole is still the enveloping surface of all the vertical lines from the points on the edge $X_{26}X_{35}$ to line $X_{13}X_{14}$, which becomes curvy due to the fact that the edge is neither parallel nor vertical to $X_{13}X_{14}$. The whole 6-facets geometry, as we expected, is also the envelope of all the vertical lines from points on the facet $X_{2,6}$ to line $X_{1,3}X_{1,4}$, but it is no longer a polytope.
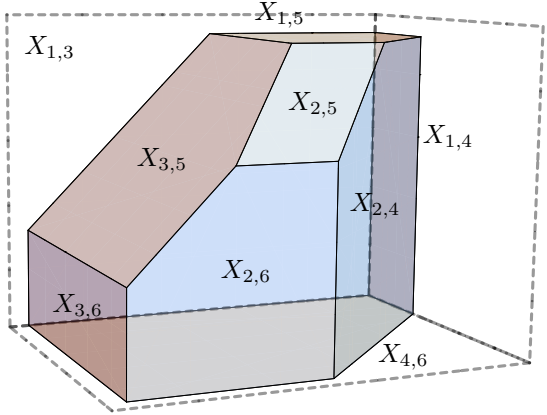
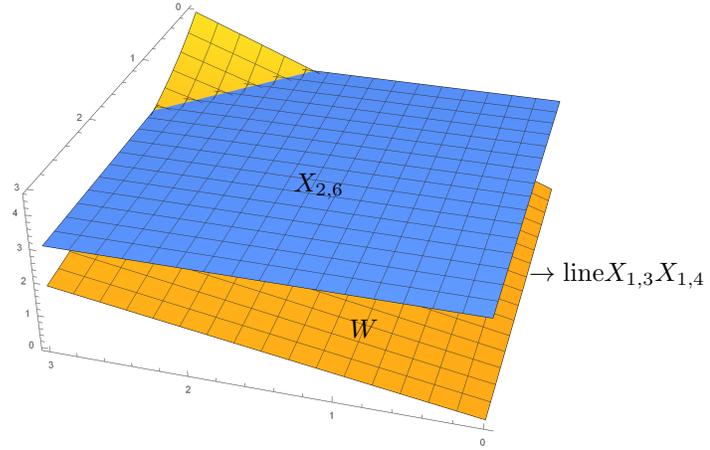
\begin{figure}[htbp]
\centering
\subfigure[$6$-point tree level amplituhedron]{
\begin{minipage}{0.4\linewidth}
\centering
\begin{overpic}[width=1.2\textwidth]
{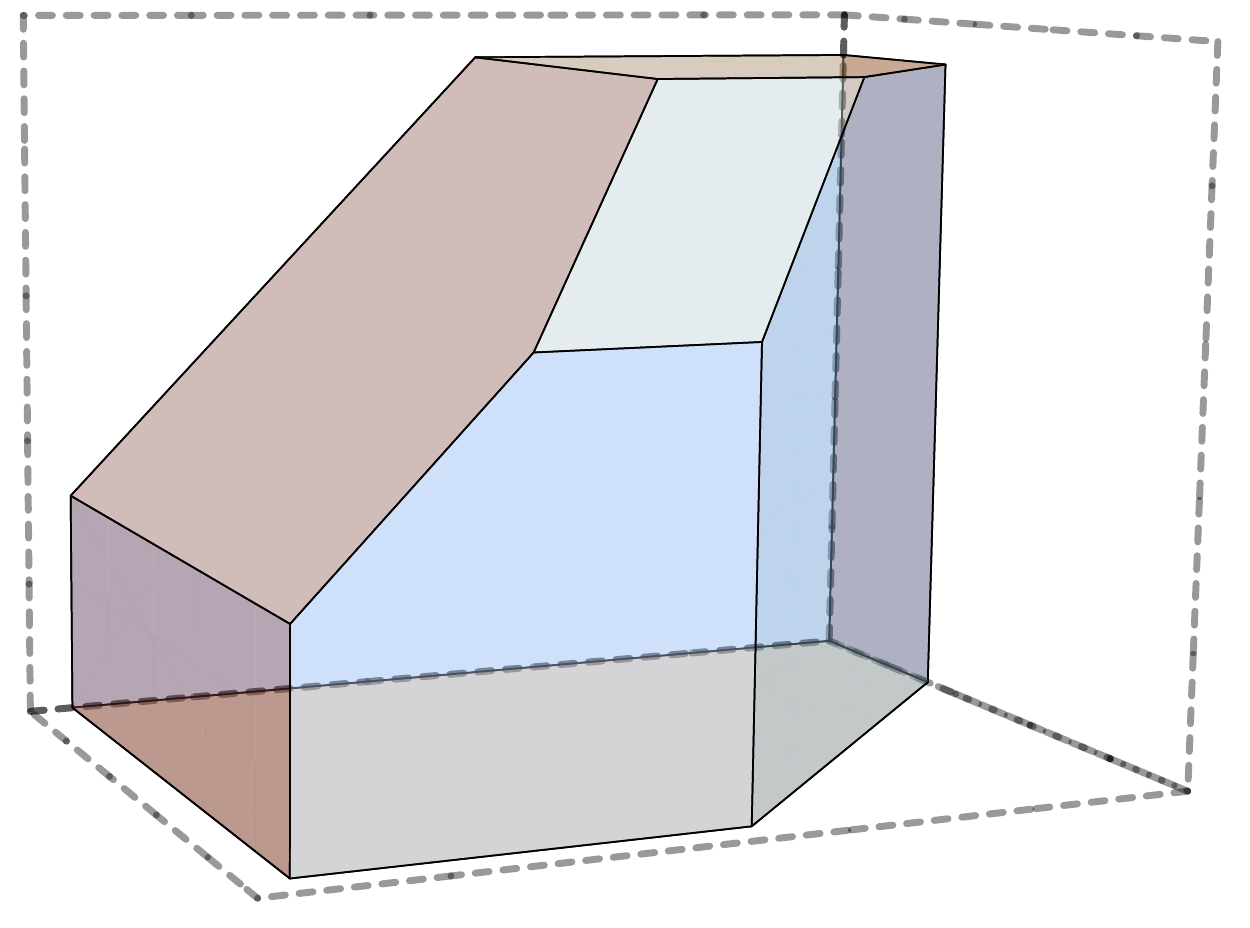}
\put(5,66){$X_{1,3}$}
\put(76,50){$X_{1,4}$}
\put(68,11){$X_{4,6}$}
\put(46,72){$X_{1,5}$}
\put(25,46){$X_{3,5}$}
\put(52,56){$X_{2,5}$}
\put(40,26){$X_{2,6}$}
\put(10,20){$X_{3,6}$}
\put(63,38){$X_{2,4}$}
\end{overpic}
\linebreak
\linebreak
\end{minipage}
}
\phantom{aaaaaaaaaaaa}
\subfigure[Spurious pole $W$ produced by rescaling $X_{1,3}$ and $X_{1,4}$]{
\begin{minipage}{0.4\linewidth}
\centering
\begin{overpic}[width=1.2\textwidth]
{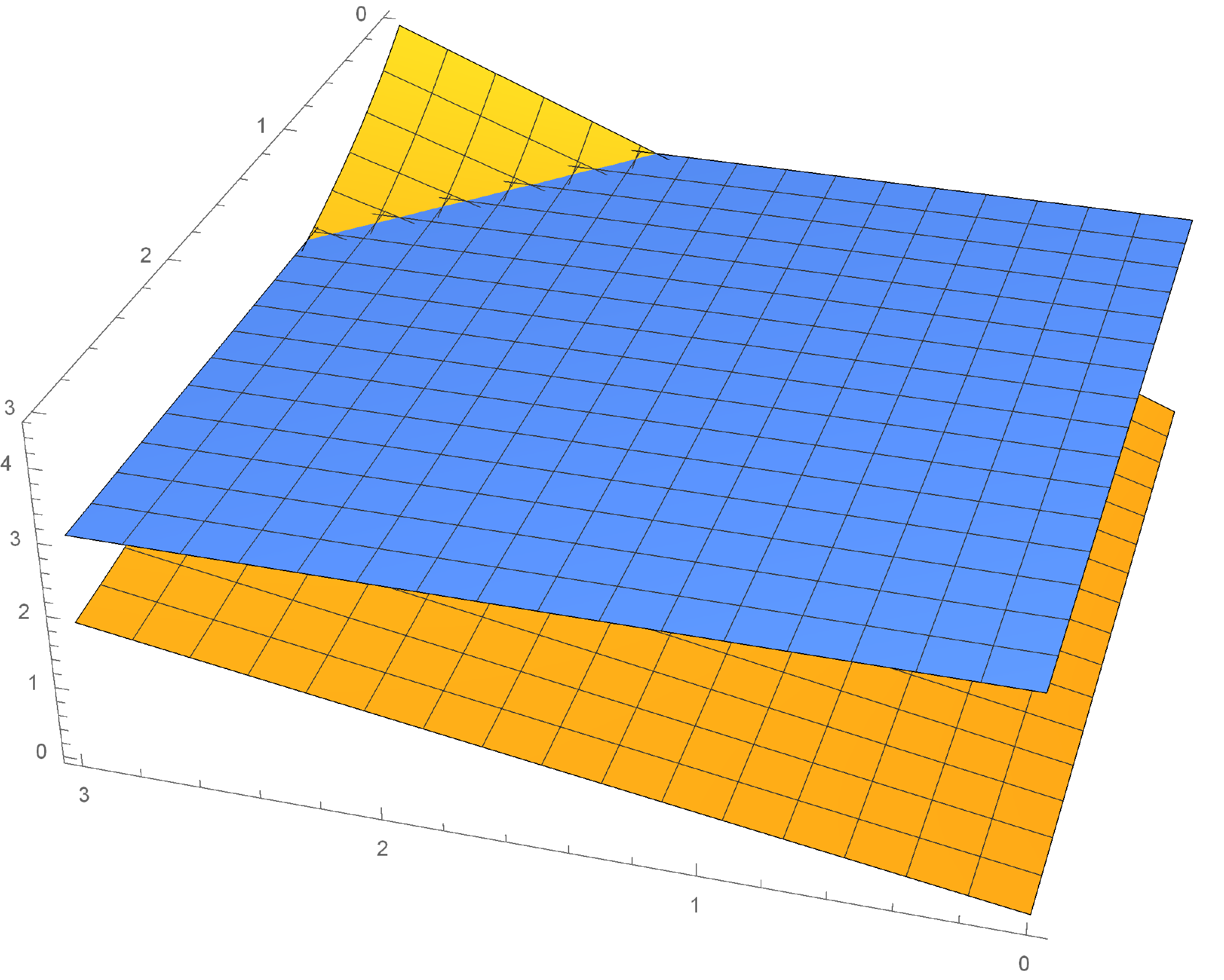}
\put(50,46){$X_{2,6}$}
\put(60,20){$W$}
\put(92,30){$\to \text{line}X_{1,3}X_{1,4}$}
\end{overpic}
\linebreak
\linebreak
\end{minipage}
}

\caption{$6$-point tree level amplituhedron and spurious pole from two variables rescaled}
{\it Two surfaces in the second figure: The blue one is the hyperplane determined by the linear function $X_{2,6}=0$, and the orange one the spurious pole, which is the envelope surface of all the vertical lines from points on the edge $X_{3,5}X_{2,6}$ onto the line $X_{1,3}X_{1,4}$.}
\end{figure}

How can such a curvy pole appear? In fact, all the spurious poles that may appear in one term are those shifted $\hat X_C(z_{B_i})$.  So for a specific $z_{B_i}\propto c_{B_i}+X_{A}$, where $c_{B_i}$ and $X_{A}$ are linear combinations of constants $C$ or undeformed variables in basis $X_{A_i}$s, certain shifted variable $X_{C}\propto z_{B_i} X_{A_j}\propto X_{A_i}X_{A_j} $, which is thus a pole of twice power in basis $X_{A_i}$. Consequently, spurious poles produced by recursion \eqref{recursion} will be at most twice power in basis, as numerators of $z_{B_i}$s and $X_{C}$s themselves are all linear functions of basis variables $X_{A_i}$.

We can then predict the geometrical interpretation of the formula $\eqref{recursion}$. For an $n$-dim generalized ABHY associahedron (can be $\mathcal{A}_{n}$, $\mathcal{\bar{D}}_n$ or others), every terms in the sum is an ``enveloping polytope" of all the vertical lines from corresponding facets to the intersection of $k$-rescaled boundaries $\{X_{A_i}\}_{i=1,\cdots,k}$ (not necessarily compatible, of course) . Generally the recursion forms a {\it curvy} triangulation of the polytope.  Certainly the triangulation can be both ``inside" or ``outside", depends on whether the images of every facets after projecting are all inside or some of them outside the polytope. When $k=1$, we simply project the polytope onto the unique facet we rescaled. While $k=n$ \eqref{recursion3}, {\it i.e.} projecting the polytope onto a facet of codimension $n$, as an $n$-codim facet is just a vertex, geometrically the recursion therefore comes back to the special version that triangulating the polytope from one point, which is the original point intersected by $n$ basis (in tree level case \eqref{recursion3} is the known formula \eqref{2}).

\section{Example at loop level}
Now let`s turn to a loop level discussion. As the second simplest case we can consider, geometrical properties of $\mathcal{\bar D}_n$ is in fact quite different from those of $\mathcal{A}_n$, as can be seen in the sections below. I will firstly offer an explicit computation for $\mathcal{\bar{D}}_4$, with an almost same rescaling as section \ref{3.1}. After that, several different recursive expressions of the integrand of $1$-loop $\phi^3$ theory amplitude will be written down.

\subsection{An explicit computation for $\mathcal{\bar{D}}_4$}
At very beginning of this section, to look deeply into the differences between polytopes $\mathcal{A}_n$ and $\mathcal{\bar{D}}_n$, let's also consider the so-called ``forgotten" limit in $\mathcal{A}_n$ case for $\mathcal{\bar{D}}_n$, {\it i.e.} apply a one-variable-rescaling recursion $X_{1,3}\to z X_{1,3}$ to $\mathcal{\bar{D}}_n$. As for every planar variables of $\mathcal{\bar{D}}_n$ we have the solution \eqref{solD} in tadpole variables, after such a rescaling, variables:
\begin{equation*}
X_3,\ \ \ X_{2,3},\ \ \ X_{3,j}\ j=1,4,\cdots,n,\ \ \& \ X_{j,2}\ j=4,\cdots,n
\end{equation*} 
will thus be shifted. So \eqref{recursion2} in this situation then reads:
\begin{equation}\label{1loop1shift}
\begin{split}
A_{1,2,\cdots,n}^{loop}=\sum_{j=5}^n(\frac1{X_{3,j}}-\frac1{X_{1,3}})\hat{A}_{3,4,\cdots,j}^{loop}\times \hat{A}_{1,2,3,j,j+1,\cdots,n}^{tree}+\sum_{j=4}^{n}(\frac1{X_{j,2}}+\frac1{X_{1,3}})\hat{A}_{2,3,\cdots,j}^{tree}\times \hat{A}_{1,2,j,j+1,\cdots,n}^{loop}\\
+(\frac1{X_{2,3}}-\frac1{X_{1,3}})\hat{A}_{1,2,3,\cdots,n}^{tree}\times \hat{A}_{2,3}^{loop}
+(\frac1{X_{3,1}}-\frac1{X_{1,3}})\hat{A}_{1,2,3}^{tree}\times \hat{A}_{1,3,\cdots,n}^{loop}+(\frac1{X_{3,4}}-\frac1{X_{1,3}})\hat{A}_{1,2,3,\cdots,n}^{tree}\times \hat{A}_{3,4}^{loop}\\
+(\frac1{X_{3}}-\frac1{X_{1,3}})\hat{A}_{1,2,3,-,+,\cdots,n}^{forward\ tree}
\end{split}
\end{equation}
Each factorized canonical function on the RHS is after the corresponding substitution \eqref{change}.

Now one can see that the same rescaling leads to quite a bit different two recursions \eqref{1loop1shift} and \eqref{tree1shift} in the two situations. Recall that in tree level case, all the physical poles emerge in such a rescaling will only be those planar variables that have an index $2$, and all the variables left unchanged form a associahedron of lower points. When it turns to the loop level, there won't be such a convenience. Although numerically they still obey the relation: number of the shifted variables, $n-3+n-3+3+1=2n-2$, is equal to the difference between numbers of planar variables of $\mathcal{\bar{D}}_n$ and $\mathcal{\bar{D}}_{n-1}$ which is $(n^2-n+1)-((n-1)^2-(n-1)+1)=2n-2$. However, unchanged planar variables in this case no longer form a $\mathcal{\bar{D}}_{n-1}$.

This difference has a strong geometrical indication. In tree level situation, all the variables left unchanged after rescaling $X_{1,3}\to zX_{1,3}$ (so that also geometrically vertical to $X_{1,3}$) are those variables without index $2$. As planar variables, they are coincident to those variables compatible with $X_{1,3}$, which combinatorially form a lower point associahedron. So that in $\mathcal{A}$ case the property is made manifest. However, this property is not held at loop level, or more specifically, when we choose tadpole variables as basis for $\mathcal{\bar{D}}_n$. For instance, the pole $X_{2,3}$ is compatible to while also linearly depends on $X_{1,3}$. Geometrically, these two facets are thus not vertical to each other for arbitrary $n$ when tadpole variables are chosen as basis.

As the first non-trivial and most easily example, let's explicitly compute the integrand of $1$-loop $4$-point amplitude from \eqref{1loop1shift}. According to the discussion above, in $n=4$ situation, we need to consider residues at $\hat X_3$, $\hat X_{31}$, $\hat X_{23}$, $\hat X_{34}$ and $\hat X_{42}$. Recursion in this case then reads:
\begin{equation}
\begin{split}
A_{1,2,3,4}^{loop}=(\frac1{X_{4,2}}+\frac1{X_{1,3}})\hat{A}_{2,3,4}^{tree}\times \hat{A}_{1,2,4}^{loop}
+(\frac1{X_{2,3}}-\frac1{X_{1,3}})\hat{A}_{1,2,3,4}^{tree}\times \hat{A}_{2,3}^{loop}
+(\frac1{X_{3,1}}-\frac1{X_{1,3}})\hat{A}_{1,2,3}^{tree}\times \hat{A}_{1,3,4}^{loop}\\+(\frac1{X_{3,4}}-\frac1{X_{1,3}})\hat{A}_{1,2,3,4}^{tree}\times \hat{A}_{3,4}^{loop}
+(\frac1{X_{3}}-\frac1{X_{1,3}})\hat{A}_{1,2,3,-,+,4}^{forward\ tree}
\end{split}
\end{equation}
All the terms except the last one can be directly written down by the indices of factorized canonical functions, or be computed from a further recursion with \eqref{change} being done before the sum in each step of the recursion. For example the second term, which is geometrically the canonical function for a direct product of a triangle and a line, together with the height factor, reads:
\begin{equation}
\Omega_{2,3}=(\frac1{X_{2,3}}-\frac1{X_{1,3}})(\frac1{X_{1,3}}+\frac1{X_{2,4}})(\frac1{(X_3-X_{2,3})X_-}+\frac1{X_2X_-}+\frac1{(X_3-X_{2,3})X_2})
\end{equation}
Replacement $X_3\to(X_3-X_{2,3})$ has been done in the third factor.

On the other hand, one may encounter difficulties when writing down the forward limit amplitudes directly from indices due to the two introduced indices $+$ and $-$. This barricade can be surmounted by making substitution of indices and corresponding planar variables, whose rule is putted in Appendix \ref{substitude}. After those procedures are done, the original forward limit amplitude reads: 

\begin{equation}\label{6forward}
\begin{split}
\frac{1}{X_1 X_2 X_{13}}+\frac{1}{X_- X_{23} X_{13}}+\frac{1}{X_2 X_{23} X_{13}}+\frac{1}{X_- X_{31} X_{13}}+\frac{1}{X_2 X_4 X_{24}}+\frac{1}{X_- X_{23} X_{24}}+\frac{1}{X_2 X_{23} X_{24}}\\
+\frac{1}{X_- X_{34} X_{24}}+\frac{1}{X_4 X_{34} X_{24}}+\frac{1}{X_4 X_1 X_{31}}+\frac{1}{X_1 X_{13} X_{31}}+\frac{1}{X_- X_{34} X_{31}}+\frac{1}{X_4 X_{34} X_{31}}+\frac{1}{X_1 X_2 X_4}
\end{split}
\end{equation} 
After the replacement
\begin{equation*}
\begin{split}
X_{23}\to X_{23}-X_{3}\\
X_{34}\to X_{34}-X_{3}\\
X_{31}\to X_{31}-X_{3}\\
X_{42}\to X_{42}+X_{3}
\end{split}
\end{equation*}
is done and height factor $(\frac1{X_{3}}-\frac1{X_{1,3}})$ is added, finally we arrive at the result of term $\Omega_{3}$ \eqref{3}.

Explicit result of this recursion is putted in the appendix. It can be seen that this recursion leads to a much more complicated computation than the tree level case.
\subsection{Projective recursions for $\mathcal{\bar{D}}_n$}
To make our computation easier, we then would like to look for a rescaling that shifts least many planar poles. Inspired by the solution \eqref{solD}, we can firstly consider the deformation:
\begin{equation}
X_-\to zX_-
\end{equation}
{\it i.e.} dividing the polytope after projecting each facet of the polytope onto the tadpole facet $X_-$. According to \eqref{solD}, only cut facets $X_i$, which linearly depend on $X_-$ contribute to the final answer. By equation \eqref{recursion2}, we can directly write down a recursion, which is a sum of $n$ terms:
\begin{equation}\label{forward}
A_{1,2,\cdots,n}^{1-loop}=\sum_{i=1}^n (\frac1{X_i}+\frac2{X_-})\hat{A}^{forward\ tree}_{1,2,\cdots,i,-,+,i+1,\cdots,n}
\end{equation}
Note that evaluating $z$ at each $z_i$ is now equal to do the replacement $X_i\to X_i-X_-$ in each factorized amplitude. Also, this result was presented in \cite{Arkani-Hamed:2019vfh} derived from ``forward limit triangulation". Geometrically, after a projection onto the tadpole facet, only those loop variables $X_{i}$ have non-trivial images, which finally contribute to the one-loop integrand. The one-loop level integrand can then be computed totally from tree level amplitudes following the factorization properties of $\bar{\mathcal{D}}_n$ on its cut facets. 

Similar to the tree level case, now a wider range of recursion relations for the  integrand of amplitude can we write down. For instance, let`s consider a second rescaling as:
\begin{equation}
X_k\to zX_k
\end{equation}
Here we need to choose a group of new basis variables. Without loss of generality, using the variables $\{X_+,X_k,X_{1,j}\}$ to represent all those $X$, one has:
\begin{equation}
\begin{split}
&X_-=2c_A+2X_{1,k}-X_+-2X_k\\
&X_i=X_k+X_{1,i}-X_{1,k}\ (i\neq k)
\end{split}
\end{equation}
and others unchanged. We therefore need to compute the contribution from poles at both $X_i$s and $X_-$. Similar procedure gives the result:
\begin{equation}
A_{1,2,\cdots,n}^{1-loop}=(\frac1{X_-}+\frac1{2X_k})\hat{A}_{1,2,\cdots,n}^{tadpole}+\sum_{i\neq k}(\frac1{X_i}-\frac1{X_k})\hat{A}^{tree\ forward}_{1,2,\cdots,i,-,+,i+1,\cdots,n}
\end{equation}
where the presence of ${A}_{1,2,\cdots,n}^{tadpole}$ is because the fact that facet $X_-$ appears to be a $\mathcal{B}_{n-1}$. Similar to the first case, this recursion is also an $n$-term sum.

At the end, we should also mention that we have a similar recursive formula for $1$-loop integrand as the one for tree level \eqref{2}, which is the standard triangulation for $\mathcal{\bar{D}}_n$. For instance, if we choose $\{X_i\}_{i=1,\cdots,n}$ as basis, {\it i.e.} triangulating the polytope from the vertex intersected by all the cut facets, the recursion \eqref{recursion3} will then read:
\begin{equation}\label{pointtri}
\begin{split}
A_{n}^{1-loop}(X,C)=\sum_{i<j-1}\frac{z_{i,j}^{n}}{X_{i,j}}A_{i,\cdots,j}^{tree}(z_{i,j}X,C)\times A_{1,\cdots,i,j,\cdots,n}^{loop}(z_{i,j}X,C)\\
+\sum_{i<j}\frac{z_{j,i}^{n}}{X_{j,i}}A_{i,\cdots,j}^{loop}(z_{j,i}X,C)\times A_{1,\cdots,i,j,\cdots,n}^{tree}(z_{j,i}X,C)
+\frac{z_-^n}{X_-}A_{n}^{tadpole}(z_-X,C)
\end{split}
\end{equation}
which is also presented in \cite{Arkani-Hamed:2019vfh}. Certainly it is the worst recursion, as after an $n$-basis rescaling it introduces most physical poles. Then the most terms we should consider ($(n-1)^2$ terms).

For example, in the case $\mathcal{\bar {D}}_4$, following the formula \eqref{forward} we can write down a four terms sum:
\begin{equation}
A_{1,2,3,4}^{loop}=(\frac1{X_1}+\frac2{X_-})A^{forward\ tree}_{1,-,+,2,3,4}+(\frac1{X_2}+\frac2{X_-})A^{forward\ tree}_{1,2,-,+,3,4}+(\frac1{X_3}+\frac2{X_-})A^{forward\ tree}_{1,2,3,-,+,4}+(\frac1{X_4}+\frac2{X_-})A^{forward\ tree}_{1,2,3,4,-,+}
\end{equation}
On the other hand, \eqref{pointtri} in this situation being a $9$-term sum, reads (under cyclic indices):
\begin{equation}
\begin{split}
A_{1,2,3,4}^{loop}=\frac{z_-^n}{X_-}A_{n}^{tadpole}(z_-X,C)+\sum_{i=1}^4\frac{z_{i,i+1}^{n}}{X_{i,i+1}}A_{i,i+1}^{loop}(z_{i,i+1}X,C)\times A_{1,\cdots,n}^{tree}(z_{i,i+1}X,C)\\
+\sum_{i=1}^4\frac{z_{i,i+2}^{n}}{X_{i,i+2}}A_{i,i+1,i+2}^{tree}(z_{i,i+2}X,C)\times A_{i+2,\cdots,i}^{loop}(z_{i,i+2}X,C)
\end{split}
\end{equation}

\section{Discussion}
In this note we generalized the recursion formula we found in \cite{He:2018svj}, which now was derived from deformation of arbitrary many planar variables $\{X_{A_i}\to zX_{A_i}\}$ in basis. After the nice fact that such a transformation will never introduce residue at infinity for the function $\frac{z^k}{z-1}A_{n}^{arb}(X,C)$, we can always rewrite tree amplitude and $1-$loop level integrand for $\phi^3$ bi-adjoint theory  as a sum of residues at several finite $z_i$s. Each one corresponds to a shifted planar variable $\hat X_{B_i}(z)=0$. Generally speaking, the complexity of the recursive computation depends on the variable we rescaled in each step. The less variables we rescale, or the less planar variables linearly dependent on the basis variables we rescale, the less terms will be introduced in the formula. Illustrating by examples, we also discovered its geometrical meaning that doing a recursive computation by the formula is equivalent to projecting the ABHY associahedron onto some lower dimensional boundaries and triangulating the polytope. Of course, the simplest recursion relation follows from a one-variable deformation, which is a projection of the whole amplituhedron onto one of its facets. Similar to the case when variables in basis are all deformed, the recursion formula is also ``BCFW" like and it can be predicted that the final result of the amplitude computed from the formula will be still a sum of products of some ``R-invariant" like terms. The only difference is that now they are not canonical functions of simplices, but several ``prism-like" geometries, which is a direct result of projective triangulation.

There are several remarks in the end of the note we should give. It should be emphasized again that although only cases $\mathcal{A}$ and $\mathcal{D}$ were discussed explicitly in this note, the recursive formulae can also be applied to generalized associahedra $\mathcal{B}$ and $\mathcal{C}$ as ABHY realizations for these two kinds of cluster polytopes are also straightforward \cite{Li:2018mnq,Arkani-Hamed:2019vfh}. Moreover, canonical functions of arbitrary polytopes that can be realized by ``C-independent" ABHY conditions can also be computed by the formulae we discussed, which have been proven in section 2.2. For instance, the new recursion formula can also be applied to Cayley polytopes, whose ABHY realization can be directly read from labelled trees \cite{Gao:2017dek}. Furthermore, the ABHY formalism was extended to tree level bi-adjoint $\phi^p$ theory, whose amplituhedron are so called ``Stokes polytopes" \cite{Banerjee:2018tun,Raman:2019utu}. It is hopeful to discover a similar recursion formula for $\phi^p$ theory like basic $\phi^3$. However, an arbitrary ABHY polytope may not satisfy the factorization properties the finite type cluster polytopes have, whose facets are still or products of lower dimensional ones. So that usually we cannot derive a recursion relation for arbitrary polytope based on a similar deformation of the planar variables in its ABHY realizations, but only represent its canonical function by canonical functions of the facets.

Also, although supported by the example, the claim that transformation of arbitrary many basis \eqref{1} and the corresponding recursion formula \eqref{recursion} indeed leads to a ``curvy triangulation" of generalized associahedra is still a conjecture. Naively one may think that, combining several positive geometries whose facets are curvy to get a polytope and its canonical function seems like making the problem more complicated. But we have seen that the canonical function of each curvy positive geometry, which is difficult to compute directly from the definition, arises naturally in the recursion formula \eqref{recursion}. Such a interesting structure may indicate some new methods to deal with the canonical function of a positive geometry with curvy facets. And this point of view may also help us in problem of $N=4$ SYM theory amplitudes and its amplituhedron. We leave this issue in future investigations. 

\section*{Acknowledgments}
The author would like to thank Song He for suggesting the project, constant discussions and patient guidances during the project. Special thanks to Nima Arkani-Hamed, Song He, Giulio Salvatori, and Hugh Thomas for sharing unpublished material \cite{Arkani-Hamed:2019vfh}. Thanks Giulio Salvatori for advice and comments he offered me after reading the draft. Also thanks my colleagues in amplitude group from ITP for advice offered in our group meetings. 
\appendix

\section{Substitution rule for the forward limit tree indices}\label{substitude}
In section 4.1 one may face trouble to write down the factorized amplitude of the forward limit, as the recursion introduces two ``cut" indices $+$ and $-$. In fact, introducing the internal indices won't cause any difficulties in some cases. For example in tree level, a factorization of tree level diagram always introduces an internal index $I$. But we can always substitute such an index by combination of some original indices after a momenta conservation law. $-$ and $+$ appearing here, on the other hand, satisfy no conservation rule like $I$ does in tree level case, which finally gives rise to the trouble for our recursion.

To conquer such a difficulty, we should either obtain the forward limit tree amplitude directly from the vertices of cut facet, or make a suitable substitution of the indices from forward limit tree to ordinary tree amplitude indices. So that we can write down the terms appearing in the forward limit directly by the vertices of the associahedron. In this section, let's find out how to directly read the planar variables appearing in forward limit tree amplitude:
\begin{equation*}
A_{n+2}^{forward\ tree}(1,2,\cdots,i,-,+,i+1,\cdots,n)
\end{equation*}

If we regard the new internal indices $-$ and $+$ also as numbers, all the planar variables in this amplitude will read:
\begin{equation*}
X_{i,j}\ \ (1\leq i<j-1\leq n), \ \ X_{k,-}\ or\ X_{-,k} \ (1\leq k\leq n\ \&\ k\neq i)\ and\ \ X_{k,+}\ or\ X_{+,k}(1\leq k\leq n\ \&\ k\neq i+1)
\end{equation*}
Now we make the substitution as below
\begin{itemize}
\item[1.]All the $X_{i,j}$s are left unchanged.
\item[2.]Replace the $X_{k,-}$ or $X_{-,k}$ with $X_k$.
\item[3.]For $1\leq k\leq i$, replace $X_{k,+}$ with $X_{ik}$. Especially, $X_{i,i-1}=X_{i-1,i}$, and $X_{i,+}$ itself is replaced by $X_-$. For $i+2\leq k\leq n$ replace $X_{+,k}$ with $X_{ki}$.
\end{itemize}
Let's check this rule by some examples. For a four point example $A_4^{forward}(1,-,+,2)$ (recall that $\mathcal{\bar{D}}_2$ is a triangle with three facets $X_-$, $X_{1}$ and $X_2$), planar variables are $X_{1,+}=X_-$ and $X_{-,2}=X_2$, which is trivially the two vertices of the facet $X_1$.

Now we move to $n=5$ case, for example $A_5^{forward}(1,2,-,+,3)$, which is the cut facet $X_2=0$ of the polytope $\mathcal{\bar{D}}_3$. Its planar variables, by the rules above, is:
\begin{equation*}
X_{1,-}=X_1,\ \ X_{1,+}=X_{1,2},\ \ X_{2,+}=X_-,\ \ X_{2,3},\ \ X_{-,3}=X_3
\end{equation*}
Moreover, their compatibility relations, which can be read from the facet structure of $\mathcal{\bar{D}}_3$, are also naturally satisfied by directly inheriting the relations of the variables before doing replacement. Therefore, the forward limit amplitude reads:
\begin{equation}\label{B1}
\begin{split}
A_5^{forward}(1,2,-,+,3)=\frac1{X_{1,-}X_{1,+}}+\frac1{X_{1,+}X_{2,+}}+\frac1{X_{2,+}X_{2,3}}+\frac1{X_{2,3}X_{-,3}}+\frac1{X_{1,-}X_{-,3}}\\
=\frac1{X_{1}X_{1,2}}+\frac1{X_{1,2}X_{-}}+\frac1{X_{-}X_{2,3}}+\frac1{X_{2,3}X_{3}}+\frac1{X_{1}X_{3}}
\end{split}
\end{equation} 
Furthermore, this substitution rule is even obeyed when we go on factorizing such a forward limit tree amplitude to lower point, which is also easy to check by the factorization property of $\mathcal{A}_n$ polytopes.

Back to the $A_{1,2,3,-,+,4}^{forward\ tree}$ case, all the variables appearing here are:
\begin{equation*}
X_{1,3},\ \ X_{1,-}=X_1,\ \ X_{1,+}=X_{3,1},\ \ X_{2,-}=X_2,\ \ X_{2,+}=X_{2,3},\ \ X_{2,4},\ \
X_{3,+}=X_-,\ \ X_{3,4},\ \ X_{-,4}=X_{4}
\end{equation*}
After a familiar recursive computation of tree level amplitude \eqref{tree1shift13}, or straightforward from the vertices of $\mathcal{A}_3$, we finally arrive at the answer:

\begin{gather}
\frac{1}{X_1 X_2 X_{13}}+\frac{1}{X_- X_{23} X_{13}}+\frac{1}{X_2 X_{23} X_{13}}+\frac{1}{X_- X_{31} X_{13}}+\frac{1}{X_2 X_4 X_{24}}+\frac{1}{X_- X_{23} X_{24}}+\frac{1}{X_2 X_{23} X_{24}}\notag\\
+\frac{1}{X_- X_{34} X_{24}}+\frac{1}{X_4 X_{34} X_{24}}+\frac{1}{X_4 X_1 X_{31}}+\frac{1}{X_1 X_{13} X_{31}}+\frac{1}{X_- X_{34} X_{31}}+\frac{1}{X_4 X_{34} X_{31}}+\frac{1}{X_1 X_2 X_4}
\end{gather}

\section{Explicit result of $A^{tree}_6$}
In this appendix we list all the terms of $6$-point tree level canonical function when rescaling $X_{13}X_{14}$. Below $\Omega_{ij}=\frac{N_{ij}}{D_{ij}}$, and $A_6=\sum_{ij}\Omega_{ij}$. Note that in this section except those physical poles we expand every planar variables on the basis $\{X_{1,i}\}_{i=3,4,5}$ as it helps the readers to figure out the curvy spurious poles the recursion produces.

Numerators:
\begin{equation}
\begin{split}
N_{24}=C_{13} (C_{14} (C_{15}+C_{25}+C_{35}) (-C_{13} X_{14}+C_{15} (X_{13}-X_{14})+X_{15} (X_{13}-X_{14}))\\
+C_{15} X_{15} (-C_{13} X_{14}+C_{15} (X_{13}-X_{14})+C_{25} (X_{13}-X_{14})+C_{35} X_{13}-C_{35} X_{14})\\
+C_{14}^2 (X_{13}-X_{14}) (C_{15}+C_{25}+C_{35}))
\end{split}
\end{equation}
\begin{equation}
N_{25}=C_{15} C_{24} (C_{13}+C_{14}+X_{15})^2
\end{equation}
\begin{equation}
\begin{split}
N_{26}=-(C_{13}+C_{14}+C_{15})^2 (-C_{15} (C_{25} (-C_{13} X_{14}+C_{24} X_{14}+C_{35} X_{13}-X_{13} X_{15}+X_{14} X_{15})\\
+C_{24} C_{35} X_{14}+C_{25}^2 X_{13})-C_{13} C_{24} C_{25} X_{14}-C_{13} C_{24} C_{35} X_{14}-C_{13} C_{25} X_{14} X_{15}\\
-C_{14} (X_{13}-X_{14}) (C_{15} C_{25}-C_{24} (C_{25}+C_{35})-C_{25} X_{15})+C_{15}^2 C_{25} (X_{14}-X_{13})+C_{24}^2 C_{25} X_{13}\\
+C_{24}^2 C_{35} X_{13}+C_{24} C_{25}^2 X_{13}+C_{24} C_{25} C_{35} X_{13}+C_{24} C_{25} X_{13} X_{15}+C_{24} C_{35} X_{13} X_{15}\\
+C_{25}^2 X_{13} X_{15}+C_{25} C_{35} X_{13} X_{15})
\end{split}
\end{equation}
\begin{equation}
\begin{split}
N_{35}=(C_{14}+C_{24}+X_{15}) (C_{13}^2 X_{14} (-(C_{15}+C_{25}))+C_{13} (C_{15}+C_{25}) (C_{14} (X_{13}-2 X_{14})\\
-C_{15} X_{14}+C_{24} X_{13}+X_{13} X_{15}-X_{14} X_{15})+C_{14}^2 (C_{15}+C_{25}) (X_{13}-X_{14})\\
+C_{14} (C_{15}^2 (-X_{14})+C_{15} (C_{24} X_{13}-C_{25} X_{14}+2 X_{13} X_{15}-X_{14} X_{15})\\
+C_{24} C_{25} X_{13}+C_{25} X_{15} (X_{13}-X_{14}))+C_{15} X_{15} (-C_{15} X_{14}+C_{24} X_{13}-C_{25} X_{14}+X_{13} X_{15}))
\end{split}
\end{equation}
\begin{equation}
N_{36}=(C_{13}+C_{14}+C_{15})(C_{14}+C_{24}+C_{15}+C_{25})C_{35}
\end{equation}
Denominators:
\begin{equation}
\begin{split}
D_{24}=X_{14} X_{15} X_{24} X_{46} (C_{13} X_{14}+C_{14} X_{14}-C_{14}X_{13}+C_{15} X_{14}-C_{15}X_{13})\\
(C_{13} X_{14}+C_{14} X_{14}-C_{14}X_{13}+X_{14} X_{15}-X_{13}X_{15})
\end{split}
\end{equation}
\begin{equation}
\begin{split}
D_{25}=X_{15} X_{25}(C_{15}-X_{15}) (-C_{13} X_{14}+C_{14} X_{13}-C_{14} X_{14}+X_{13} X_{15}-X_{14} X_{15}) \\
(-C_{13} X_{14}+C_{14} X_{13}-C_{14} X_{14}+C_{24} X_{13}+X_{13} X_{15}-X_{14} X_{15})
\end{split}
\end{equation}
\begin{equation}
\begin{split}
D_{26}=(X_{26}X_{46}(-C_{14} X_{13} - C_{15} X_{13} + C_{13} X_{14} + C_{14} X_{14} + C_{15} X_{14})(C_{15} - X_{15})\\
(-C_{14} X_{13} - C_{15} X_{13} - C_{24} X_{13} - C_{25} X_{13} + C_{13} X_{14} + C_{14} X_{14} + C_{15} X_{14})\\ (-C_{14} X_{13} - C_{24} X_{13} + C_{13} X_{14} + C_{14} X_{14} + C_{15} X_{14} - X_{13} X_{15})
\end{split}
\end{equation}
\begin{equation}
\begin{split}
D_{35}=X_{13} X_{15}X_{35} (C_{15}+C_{25}-X_{15})  (-C_{13} X_{14}+C_{14} X_{13}-C_{14}X_{14}-C_{15} X_{14}+C_{24} X_{13}+X_{13} X_{15})\\(-C_{13} X_{14}+C_{14} X_{13}-C_{14} X_{14}+C_{24} X_{13}+X_{13} X_{15}-X_{14} X_{15})
\end{split}
\end{equation}
\begin{equation}
D_{36}=(X_{13}X_{36}X_{46}(C_{15}+C_{25}-X_{15})(C_{24}X_{13} + C_{25} X_{13} + C_{14} X_{13} - C_{14}X_{14} + C_{15} X_{13} -C_{15} X_{14} - C_{13} X_{14})
\end{equation}
\section{Explicit result of $A^{1-loop}_4$}
We list the result of $A^{1-loop}_4$ in this section. Every terms are represented by planar variables $X_A$s.
\begin{equation}\label{3}
\begin{split}
\Omega_3=(X_- (X_1 (X_2 (X_3-X_{13}) (X_3-X_{23}) (X_3-X_{24}-X_{31})-((X_3-X_{13}) (X_3-X_4-X_{23})+X_4 X_{24}) \\
(X_3-X_{31}) (X_3-X_{34}))+(X_3-X_4-X_{13}) (X_3-X_{23}) X_{24} (X_2-X_3+X_{31}) (X_3-X_{34}))\\
+X_1 X_2 X_4 (2 X_3^3-(2 X_{13}+X_{23}+3 X_{24}+2 X_{31}+X_{34}) X_3^2+(X_{24} X_{31}+X_{23} (2 X_{24}+X_{31})\\
+(2 X_{24}+X_{31}) X_{34}+X_{13} (X_{23}+X_{24}+2 X_{31}+X_{34}))\\
X_3-X_{24} (X_{23}+X_{31}) X_{34}-X_{13} (X_{23} (X_{24}+X_{31})+X_{31} X_{34})))\\
/(X_- X_1 X_2 X_3 X_4 X_{13} (X_3-X_{23}) X_{24} (X_3-X_{31}) (X_3-X_{34}))
\end{split}
\end{equation}
\begin{equation}
\begin{split}
\Omega_{31}=((X_1+X_4) (X_3-X_{31}) (X_{13}-X_{31}) (X_{31}-X_{34})-(X_1 (X_3 (X_{13}-X_{31})+X_{31} (X_{31}-X_{13})\\
+X_4 (X_{13}-2 X_{31}+X_{34}))-X_4 (X_3-X_{31}) (X_{31}-X_{34})) X_{41}+X_- ((X_3-X_{31}) (X_{13}-X_{31}) (X_{31}-X_{34})\\
+(X_1 (X_{31}-X_{13})+(X_4+X_{13}-X_{31}) (X_{31}-X_{34})) X_{41}))\\
/(X_- X_1 X_4 X_{13} (X_3-X_{31}) X_{31} (X_{31}-X_{34}) X_{41})
\end{split}
\end{equation}
\begin{equation}
\Omega_{23}=\frac{\left(X_-+X_2+X_3-X_{23}\right) \left(X_{13}-X_{23}+X_{24}\right)}{X_- X_2 X_{13} \left(X_3-X_{23}\right) X_{23} X_{24}}
\end{equation}
\begin{equation}
\Omega_{34}=\frac{\left(X_-+X_3+X_4-X_{34}\right) \left(X_{13}-X_{34}\right) \left(X_{24}+X_{31}-X_{34}\right)}{X_- X_4 X_{13} X_{24} X_{34} \left(X_{34}-X_3\right) \left(X_{34}-X_{31}\right)}
\end{equation}
\begin{equation}
\begin{split}
\Omega_{42}=((X_2 X_4 X_{24} (X_{12}+X_{41})+X_- (X_2 X_{12} X_{24}+(X_1 X_{12}+(X_4+X_{12}) X_{24}) X_{41})+X_1 (X_4 (X_{12}+X_{24}) X_{41}\\
+X_2 X_{12} (X_{24}+X_{41}))) (X_{13}+X_{42}))/(X_- X_1 X_2 X_4 X_{12} X_{13} X_{24} X_{41} X_{42})
\end{split}
\end{equation}
the final result then reads:
\begin{equation}
\Omega_4^{loop}=\Omega_{3}+\Omega_{31}+\Omega_{23}+\Omega_{34}+\Omega_{42}
\end{equation}

\bibliographystyle{utphys}
\bibliography{refsout}
\end{document}